\documentclass[preprint2]{aastex}

\shorttitle{The apparent proper motion and the VLBI structure of compact radio sources}

\shortauthors{Mo\'or et al.}

\begin{document}


\title{On the connection of the apparent proper motion and the VLBI structure of compact radio sources}

\author{A.~Mo\'or\altaffilmark{1,2,3}}
\email{moor@konkoly.hu}
\author{S.~Frey\altaffilmark{1,2}}
\email{frey@sgo.fomi.hu}
\author{S.B.~Lambert\altaffilmark{4}}
\email{sebastien.lambert@obspm.fr}
\author{O.A.~Titov\altaffilmark{5}}
\email{oleg.titov@ga.gov.au}
\author{J.~Bakos\altaffilmark{6,7,8}}
\email{jbakos@iac.es}
\altaffiltext{1}{F\"OMI Satellite Geodetic Observatory, P.O. Box 585, H-1592 Budapest, Hungary}
\altaffiltext{2}{MTA Research Group for Physical Geodesy and Geodynamics, P.O. Box 91, H-1521 Budapest, Hungary}
\altaffiltext{3}{Konkoly Observatory of the Hungarian Academy of Sciences, P.O. Box 67, H-1525 Budapest, Hungary}
\altaffiltext{4}{Observatoire de Paris, D\'epartement Syst\`emes de R\'ef\'erence Temps Espace (SYRTE), CNRS/UMR8630, 75014 Paris, France}
\altaffiltext{5}{Geoscience Australia, P.O. Box 378, Canberra 2601, Australia}
\altaffiltext{6}{Instituto de Astrof\'isica de Canarias, E-38205 La Laguna, Tenerife, Spain}
\altaffiltext{7}{Departamento de Astrof\'isica, Universidad de La Laguna, E-38205 La Laguna, Tenerife, Spain}
\altaffiltext{8}{Department of Astronomy, Lor\'and E\"otv\"os University of Sciences, H-1518, Budapest, Hungary}


\begin{abstract}

Many of the compact extragalactic radio sources that are used as fiducial points to define
the celestial reference frame are known to have proper motions detectable with long-term 
geodetic/astrometric Very Long Baseline Interferometry (VLBI) measurements. 
These changes can be as high as several hundred micro-arcseconds per year for certain objects. 
When imaged with VLBI at milli-arcsecond (mas) angular resolution, these sources (radio-loud active galactic nuclei) 
typically show structures dominated by a compact, often unresolved ``core'' and a one-sided ``jet''.

The positional instability of compact radio sources is believed to be connected with changes 
in their brightness distribution structure. For the first time, we test this assumption in a statistical sense on a large sample,
not only for individual objects. We investigate a sample of 62 radio sources for which reliable long-term time series of 
astrometric positions as well as detailed 8-GHz VLBI brightness distribution models are 
available. We compare the characteristic direction of their extended jet structure and the 
direction of their apparent proper motion.

We present our data and analysis method, and conclude that there is indeed a correlation between the two 
characteristic directions. However, there are cases where the $\sim$$1-10$-mas scale VLBI jet directions are significantly 
misaligned with respect to the apparent proper motion direction. 

\end{abstract}





\section{Introduction} \label{intro}

The International Celestial Reference Frame (ICRF) is a quasi-inertial
reference frame defined through the accurate radio positions of extragalactic
sources (active nuclei of quasars, radio galaxies, and BL Lac objects) distributed over the sky.
Its first realization \citep[ICRF1;][]{ma98} was based on the positions of 
212 defining radio sources, using Very Long
Baseline Interferometry (VLBI) measurements made between 1979 and 1995. 

The continuous accumulation of new high-quality VLBI data, and technical, observing and data analysis
improvements made it possible to define the ICRF2 \citep{icrf2} which was adopted by
the International Astronomical Union (IAU) in 2009 as the current official realization of the 
celestial reference system. There are 295 defining sources selected for ICRF2, based on their
positional stability and compact radio structure, as deduced from nearly three decades of VLBI 
monitoring. The ICRF2 catalog lists the coordinates of 
a total of 3414 extragalactic radio sources. The new solution is aligned with the ICRF1
to ensure continuity. The new reference frame contains the positions of more than 5 times as many sources, 
better distributed over the northern and southern hemispheres. 
Its accuracy is at least 5 times better than that of the ICRF1.

Due to the frequent observations in the framework of geodetic/astrometric VLBI programs, the positional accuracy of many
of the reference sources is better than $\sim$$0.2$ milli-arcseconds (mas) in both equatorial coordinates.
It has long been known, however, that several of these sources show significant 
apparent proper motion on a scale up to 100 micro-arcseconds ($\mu$as) per year \citep[e.g. ][]{feissel-vernier03}.
These systematic
linear or non-linear variations in the position of some reference sources
can lead to inaccuracies in the
estimates of other parameters \citep[e.g. Earth orientation parameters;][and references therein]{macmillan07}.

The typical angular scale of $\sim$$1-10$\,mas of the emitting
regions in these radio sources
that can be studied with VLBI imaging at centimeter wavelengths
is significantly larger than the
formal precision of the position determination.
Variations in the brightness distribution of quasars and radio galaxies are
not rare events. It
is a general assumption that in most cases the apparent proper motions
are mainly due
to this intrinsic structural evolution,
e.g. motions of relativistic jet components away from the innermost regions of active
galactic nuclei (AGNs). This hypothesis was confirmed for some individual objects.
For example, a detailed analysis of 0923+392 (4C\,+39.25) showed that the apparent proper motion of the source
is related to structure variations \citep{fey97}. The apparent proper motion of 2200+420 (BL Lac)
was found to significantly decrease after structural corrections derived from
source maps were applied to the data set \citep{charlot02}.
In some cases, the proper motions of the quasars are not easily modeled by a linear function \citep{titov07}.
These investigations implied that the positional instabilities
can be explained by examining the variations of the source structure in these individual cases. However,
the general hypothesis has not been evaluated on a larger sample yet.

If this model can be applied for most of the sources, then we can expect
some correlation between the direction of the position change, and the
typical direction of the source ``core--jet'' structure in the sky. The
mas-scale radio structure of AGNs is dominantly one-sided. This is a
consequence of relativistic beaming which enhances the radiation
originating from the relativistic plasma outflow pointing towards the
observer, close to the line of sight. At the same time, the receding
side of the intrinsically symmetric jet is de-boosted and becomes
practically invisible. The compact, often unresolved, flat-spectrum
``core'' one can usually see in VLBI images is in fact the base of the
inner jet, characterized by an optical depth of $\tau=1$. The absolute
position of this ``core'' therefore depends on the observing frequency
\citep[e.g. ][]{blandford79,lobanov98,kovalev08}.
Even at a fixed frequency, positional changes are expected with time due
to changes in the intrinsic physical (e.g. jet Lorentz factor, electron
density) or geometric (viewing angle, jet precession) conditions within
the innermost few parsecs of the AGN environments. These changes manifest
themselves in brightness, spectral or polarisation variations,
appearance, disappearance, enhancement or fading of optically thin VLBI
``jet'' components. These phenomena are detectable with high-resolution VLBI
imaging typically over time scales of years or longer. Propagation
effects (e.g. interstellar scattering) extrinsic to the compact sources
could also lead to apparent structural or brightness modifications, down
to sub-daily time scales \citep[e.g. ][]{wagner95}. The changes could
in principle affect the absolute position of the radio centroid of a
radio-loud AGN which serves as a reference point in astrometric and
geodetic VLBI observations.

Some systematic effects could contribute to the total magnitude of the apparent 
proper motion as estimated from geodetic VLBI data. One of them is the secular aberration drift 
caused by the Galactocentric acceleration of the Solar system
barycenter. 
This acceleration $a=V^2/R$ (where $V$ is the orbital
velocity due to rotation of the Solar system around the Galactic center, and $R$ is the Galactocentric distance) estimated as 
$2.7\times10^{-13}$\,km\,s$^{-2}$ \citep[$V$ = 254\,km\,s$^{-1}$, $R$ = 8.4\,kpc,][]{reid2009} results in 
a secular aberration drift equal to 5.4
$\mu$as/year. The effect of noncircular motion was found negligible ($<$1 $\mu$as/year) by \citet{kopeikin2006}.

\citet{aberration} detected the secular aberration drift in radio source
proper motions as derived from 30 years of the geodetic VLBI observations. They found
6.4$\pm$1.5 $\mu$as/year, which is consistent with the theoretical prediction.
Thus, this effect should be removed from the apparent proper motion of individual radio sources for
a fair comparison with astrophysical data.

In the present work, we compared the characteristic position angle of
the core--jet structure of 62 radio sources derived from their
multi-epoch VLBI structure maps
with the direction of their apparent proper motion
based on the coordinate time series of regular geodetic VLBI
observations. The individual proper motions have been corrected for the secular
aberration due to the rotation of the Solar system barycenter around the center of the Galaxy.
The data sets and the data analysis are described in \S~\ref{datasets}. The results are presented in
\S~\ref{results}. Implications of our results are discussed in \S~\ref{discussion}.

\section{Data} \label{datasets}

\subsection{Description of the apparent proper motion data} \label{descdatasets1}

To derive the apparent proper motion of radio sources, we analyzed more than 5,000 geodetic 
VLBI sessions of the International VLBI Service for Geodesy and Astrometry \citep[IVS,][]{schluter2007}
data base since 1979 using the Calc~10.0/Solve~2010.05.21 geodetic VLBI
analysis software package, developed and maintained at the NASA Goddard Space Flight Center. A more detailed description 
of the analysis configuration is given in \citet{aberration}. Earth orientation parameters, station and radio 
source coordinates were estimated once per session. Tropospheric zenith delays and gradients were estimated at a higher rate. 
We used a loose constraint to tie the celestial reference frame to the ICRF2 \citep{icrf2} in order to prevent a 
spurious net rotation of a set of radio sources that could result in a fake orientation of the proper motion.

The above analysis configuration allowed us to obtain coordinate time series for all the observed radio sources. In these series, data
points resulting from less than three good observations within a session were removed and outliers were eliminated so that the reduced 
$\chi^2$ is
reasonably close to unity. Proper motions were then estimated for time series containing at least 10 points and longer than 10~years by
weighted least-squares with weights taken as the inverse of the squared formal error. 
The final sample contains the proper motions of 593 sources.

The proper motions were freed from the secular aberration drift caused by the rotation of the Sun around the Galactic center. The
contribution is given by Eqs.~(3)--(4) of \citet{aberration} wherein the coefficients $d_i$ correspond to a vector of amplitude of
6.4~$\mu$as/year directed toward $\alpha=263^{\circ}$ and $\delta=-20^{\circ}$ ($d_1=-0.7$, $d_2=-5.9$, $d_3=-2.2$).
 
\subsection{Description of the VLBI structure data} \label{descdatasets2}

In order to determine the characteristic position angle of the
core--jet structures in a large number
of radio sources, and to monitor the variability of the typical jet directions,
we used the database compiled by \citet[][hereafter P07]{piner}. This data set
is based on the US Naval Observatory's Radio Reference Frame Image Database, an ongoing program
to image radio reference frame sources regularly.
\citet{piner} used the data series of 77 sources obtained at 8\,GHz. They
fitted Gaussian brightness distribution model components interactively to the visibility data on a source-by-source basis with the
Caltech Difmap software \citep{shep94}. Their final database (see Table\,3 in P07)
includes the main parameters of the fitted components (e.g. the polar coordinates of the centers, the sizes and orientations of the major and minor axes,
and the flux densities of the Gaussians) for each source at different epochs.
All of the 77 sources in P07 can be found in the proper motion database as well. We note that 47 of them are ICRF2 defining
sources, and thus are among the most stable and less prominently extended objects.

\subsection{Data analysis} \label{dataanalysis}

\begin{figure*} 
\epsscale{1.5}
\plotone{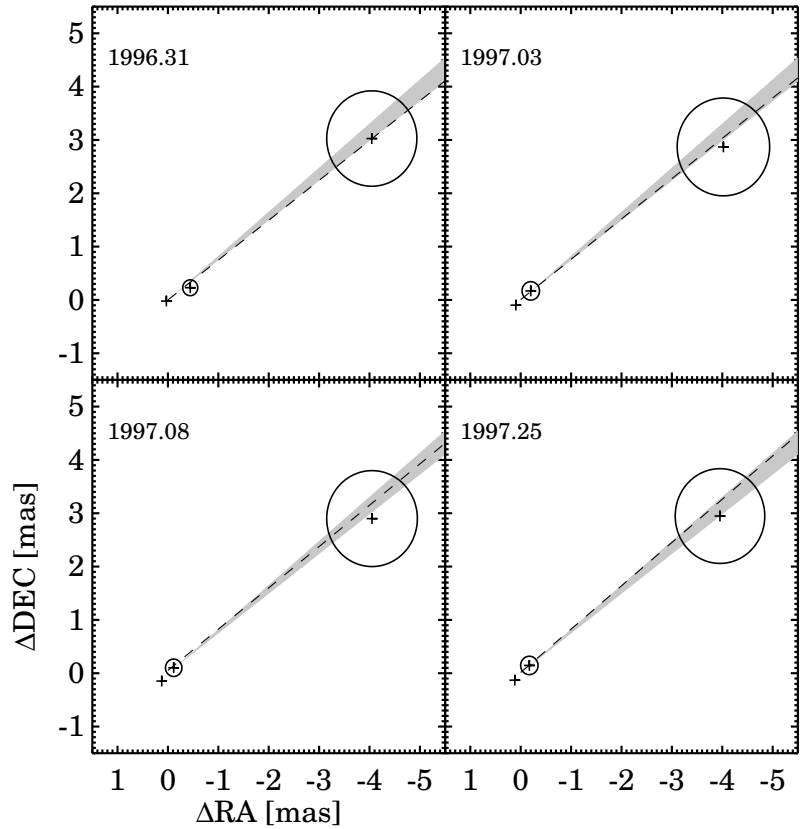}
\caption{Structure maps represented by fitted Gaussian models for 0202+149 at four
different epochs, taken from P07.
The different components of the radio source are denoted by circles, whose diameters correspond to the
full width at half maximum (FWHM) size of the fitted Gaussians.
The dashed lines show the results of our line fittings, while the grey areas correspond to the $\phi\pm\sigma_{\phi}$
region (derived from the position observations at 13 different epochs). 
\label{jetfit}}
\end{figure*}

For the selected sources, the P07 database provides model brightness distributions at 3--19 different
epochs over the years 1994--1998.
We fitted a straight line to the positions of the components that
describe a specific source using a weighted least-squares method.
A position angle was calculated
independently at each epoch.
The fit was weighted by the flux densities of the individual emission features.
Model components with ID=99 (Table~3 in P07) were not considered in our analysis, since these components could not be directly identified
with model components that appear at other epochs in P07.
The final characteristic core--jet direction ($\phi$) and its uncertainty ($\sigma_{\phi}$) were computed
as the average and the standard deviation of the position angles obtained at different epochs.
Fig.~\ref{jetfit} illustrates our fitting method for 0202+149.

The apparent proper motion values in right ascension and declination, and their uncertainties were transformed into
polar coordinates, yielding four parameters for each source: the radial proper motion ($r$) and the position angle ($\psi$) with their
uncertainties ($\sigma_r$, $\sigma_\psi$). Sources with negligible apparent proper motions (${r}/{\sigma_r} < 1$) were discarded from our
sample and were not used in the further analysis. 
The derived parameters of the final selection of 62 sources are summarized in
Table~\ref{table1}.

\section{Results}  \label{results}

The variations present in the apparent positions of AGNs are believed to be mainly caused by
changes in the brightness distribution of the sources.
According to this general picture, the direction of the apparent proper motion of a specific source should
show correlation with the typical direction of the source core--jet structure in the sky.
The multi-epoch VLBI structure maps offer a good opportunity to check whether the typical core--jet directions
are stable in time for our sample.
Both the real variations in the core--jet direction and the uncertainties of the fitted Gaussian models can contribute
to the uncertainties of the derived position angles. For most of our sources, these values are relatively low ($<$20$\degr$), implying
that the role of real variations may be negligible.
Therefore in the further analysis we assumed that the derived directions are valid
on longer time scales as well. The time span of the geodetic VLBI observations is up to five times longer than that of the structure maps analyzed in P07.

To verify our assumption, we studied an extreme example, the source OJ\,287 (0851+202) in more detail. 
The blazar OJ\,287 is known to have a rapidly changing inner 
radio jet structure \citep[e.g.][and references therein]{tate04} of which the kinematics could be revealed by frequent VLBI monitoring.
We investigated the mas-scale jet direction in OJ\,287, as defined by the component positions in 8-GHz VLBI imaging data, 
over a time interval much longer than covered by P07. To this end, we collected data from the literature (1985--1988, 1990--1996), 
and analyzed archival geodetic/astrometric VLBI visibility data (1994--2008). 
The total time range spanned is 24 years, comparable to the time range of the data from which the proper motion is estimated (\S~\ref{descdatasets1}). 

A total of 59 epochs of 8-GHz Very Long Baseline Array (VLBA) data sets were taken from the 
VLBI database of compact radio sources\footnote{\url {http://astrogeo.org/vlbi\_images/}}, from 1994 to 2008. 
In comparison, P07 used data from 15 of these epochs, between 1994 and 1998. With the Difmap software, 
we model-fitted the calibrated interferometric visibilities at 40 additional epochs where sufficient data were available, 
adopting the same method used by P07.
In addition, to extend the covered time interval backwards, we refer to \citet{vicente96} who analyzed 8-GHz 
geodetic VLBI observations at 10 epochs (jet component was detected in nine epochs) between 1985 and 1988. 
They explain the observed radio structure with a 
helical jet arising from a supermassive binary black hole system. A similar study by \citet{tate99} used 27 
epochs between 1990 and 1996. 

\begin{figure*} 
\epsscale{2.0}
\plotone{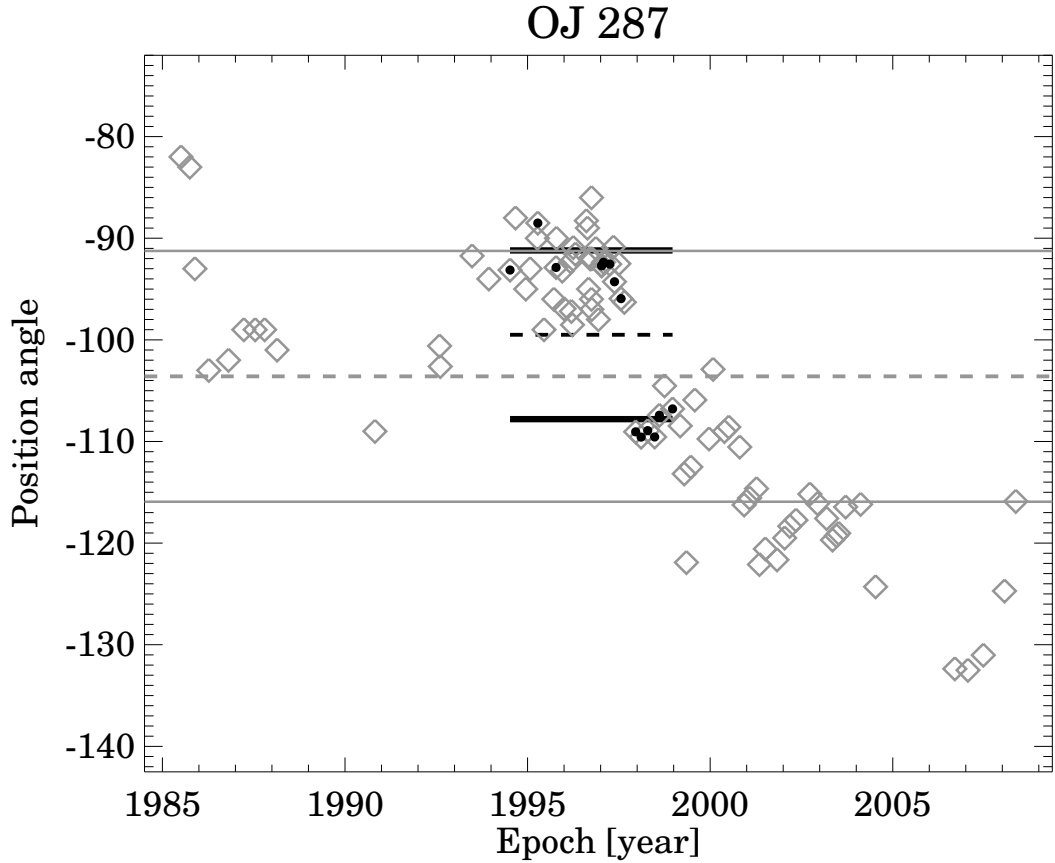}
\caption{The characteristic position angle of 0851+202 (OJ\,287) core--jet structure as a function of time (gray diamonds), 
      as derived from 8-GHz VLBI imaging data taken between 1986 and 2008. 
      The data from \citet{piner} used for the analysis presented in this paper (filled circles) 
      cover a much shorter time range (1994--1998). The source is known to have a precessing radio 
      jet which explains the clear directional variations seen in this plot. Even for this source with a highly 
      and systematically variable structure, the average position angle determined from the \citet{piner} data 
      ($-99\fdg5 \pm 8\fdg3$)
      and from the data taken over more than two decades 
      ($-103\fdg6 \pm 12\fdg3$) 
      are similar within the uncertainties. \label{oj287}}
\end{figure*}

Fig.~\ref{oj287} shows the characteristic jet position angle of OJ\,287 as a function of time. 
There is indeed a clear systematic change in this angle, quite consistent with the predictions of the 
ballistic precessing jet model of \citet{tate04}. However, from the point of view of the present study, 
it is important to note that the jet always extends nearly to the west-southwest (position angles between 
about $-80\degr$ and $-135\degr$). Within the uncertainties, the characteristic core--jet direction ($\phi$) 
for OJ\,287 we determined from the P07 data (\S~\ref{dataanalysis}) agrees with the average value over 24 
years. In other words, our method of using only the 5-year P07 data is justified since it does not introduce a 
significant bias in the characteristic jet direction, even in the case of this peculiar radio source.

The changes in the brightness distribution of the radio sources could cause
apparent positional variations in the opposite directions along the core--jet axis.
In order to define a modulo 90$\degr$ differential angle between the the two directions, 
we computed a $\phi^{'}$ angle for
each object using the following formulae: ($i$) $\phi^{'} = \phi$, if the difference between
 $\phi$ and $\psi$ is $\leq$90$\degr$; ($ii$) $\phi^{'} = \phi + 180\degr$, if the angle between
 $\phi$ and $\psi$ is $>$90$\degr$ and $\phi<0\degr$; ($iii$) $\phi^{'} = \phi - 180\degr$,
if the angle between $\phi$ and $\psi$ is $>$90$\degr$ and $\phi>0\degr$.
The differential angle ($\Delta \Phi = {|\psi - \phi^{'}|}$) is therefore the smallest angle
between the direction of the apparent proper motion and the axis
defined by the core--jet structure.
The derived $\Delta \Phi$ parameters and their $\sigma_{\Delta \Phi}$ uncertainties are listed
in the last column of Table~\ref{table1}.

\subsection{General association between the characteristic directions}

Since $\psi$ and $\phi^{'}$ are circular variables, the analysis of the possible association between
them requires special methods. In order to test the hypothesis whether the two variables are
independent, we performed a general test, following the procedures described in 
\citet{fisher95}. As a first step of this analyis, $\psi$ and $\phi^{'}$ were re-ordered to get data
pairs of ($\psi_{1}^{*},\phi_{1}^{'*}$),...,($\psi_{n}^{*},\phi_{n}^{'*}$), where 
$\psi_{1}^{*},...,\psi_{n}^{*}$ are in cyclic order, and $S_{1},...,S_{n}$ were determined as the rank
of the corresponding $\phi^{'*}$ values. Then we calculated $$\gamma_n^2 = (1/n^4)\sum_{i=0}^n
\sum_{j=0}^n (T_{i,i} + T_{j,j} - T_{i,j} - T_{j,i})^2,$$ where $$T_{i,j} = n \cdot {\rm min}(i,S_j) -
i \cdot S_j, ~~~i,j=1,...,n.$$ The total number of radio sources in the analysis was $n=62$. The null
hypothesis that $\psi$ and $\phi^{'}$ are independent can be rejected if $\gamma_n^2$ is too large. In
order to determine a critical value for $\gamma_n^2$, we performed a randomisation test 
\citep[see also ][]{fisher95}: ($a$) we added uniformly-distributed random numbers in the range of $0\degr$
and $360\degr$ to the position angles derived for the core--jet structures ($\phi^{'}$) and then
re-calculated the value of $\gamma_n^2$; ($b$) this treatment was repeated 1000 times, and the
$\gamma_n^2$ values obtained were compared with the original one. We found that the value of $\gamma_n^2$ 
for our original sample is quite high: the test showed that the null hypothesis, that the two variables are
independent, can be rejected at a level of 98.3\%.

\begin{figure*} 
\epsscale{2.2}
\plotone{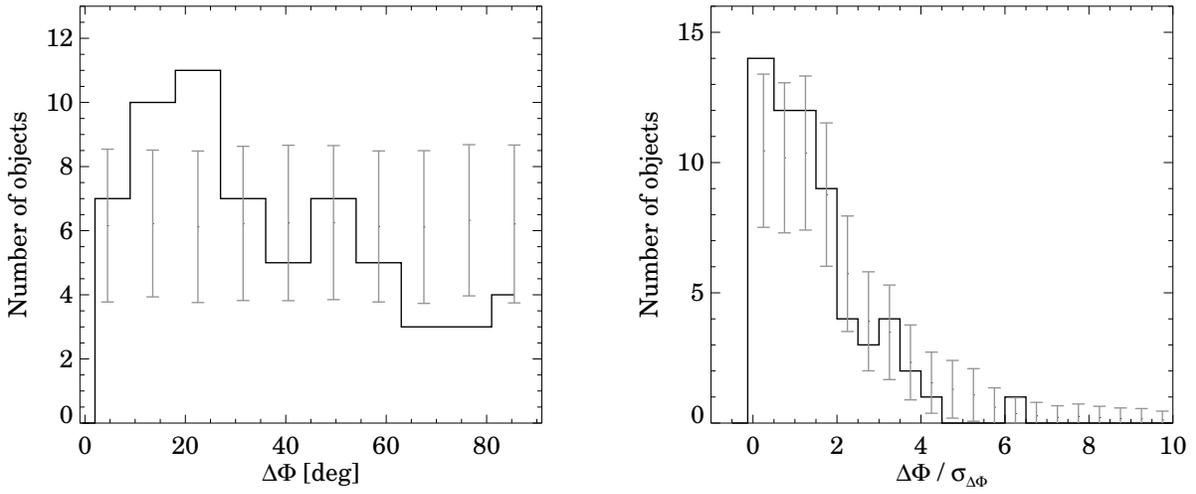}
\caption{Distribution of the $\Delta \Phi$ (left panel) and the $\frac{\Delta \Phi}{\sigma_{\Delta \Phi}}$ (right panel) values for the 62 radio sources analyzed (histograms).
Average values obtained from 1000 similar but simulated random samples are shown with vertical bars whose size indicates the standard deviation. 
\label{simulation}}
\end{figure*}

The distribution of the $\Delta \Phi$ values is presented in Fig.~\ref{simulation} (left). The histogram
shows a peak at low differential angles, which could be expected if there is correlation between
the two directions. Using the results of the previous randomisation test, we also compiled the
simulated distribution of ${\Delta \Phi}$ (Fig.~\ref{simulation}, left) and ${\Delta \Phi}/{\sigma_{\Delta
\Phi}}$ (Fig.~\ref{simulation}, right) by averaging the results of the 1000 different runs in each bin. As
Fig.~\ref{simulation} indicates, both distributions show an excess compared to the simulated distributions
at their lower end. This finding also implies that there is an association between the two characteristic directions. 
For some sources this association seems to be very strong. The best examples were selected based on the following criteria:
the differential angle between the two typical directions is $\Delta
\Phi< 5\degr$, and ${\Delta \Phi}/{\sigma_{\Delta \Phi}} < 1$. Two objects fall into this category, \object[0919-260]{0919-260}
and \object[1156+295]{1156+295}.

On the other hand, there are a couple of sources where the angle between the direction of the apparent proper motion and the axis
defined by the core--jet structure is quite large. \object[0851+202]{0851+202} (OJ\,287) is especially interesting from this point of view, 
since this object forms a clear outlier at ${\Delta \Phi}/{\sigma_{\Delta \Phi}} = 6.4$ in Fig.~\ref{simulation} (right) showing that its 
nearly orthogonal misalignment (see Table~\ref{table1}) is highly significant. 

We note that the omission of the correction for the secular aberration
in the calculation of proper motion data would not alter our
main conclusion concerning the association between the two characteristic
directions. To test this, we repeated the analysis using a data set derived 
without this correction. In that case, the null
hypothesis could be rejected at a level of 99.5\%.
Before 1990, the general deficiency of the VLBI networks, including the number
of observed sources and observing antennas per session, makes the VLBI products
less reliable \citep[see, e.g.][who reported interesting statistical results
and remarks about the VLBI evolution over the last
two decades]{gontier01,malkin2004,feissel-vernier2004,lambert09}.
In order to check the possible effect of these deficiencies on
our findings, we repeated our analysis using a recompilation of the proper motion
database without the geodetic data measured before 1990. We obtained a significance level of 96.6\% 
for the rejection of the null hypothesis. It is in good agreement with the value derived utilizing the original
data set.

\section{Discussion} \label{discussion}

Based on our sample of 62 radio-loud AGNs, there is correlation between the direction of
their apparent proper motion and the typical direction of their core--jet structure seen in
cm-wavelength VLBI images. However, this does not necessarily mean that the proper motion is only
influenced by changes in the source brightness structure. There exist individual objects where 
the apparent proper motion direction and the characteristic core--jet structure direction 
differ significantly.

In fact different angular scales are
probed by the 8-GHz VLBI imaging and the astrometric position measurements. While the typical angular
size of the core--jet structures we investigated is $\sim$$1-10$\,mas, the measured proper motion
values are at most in the $\sim$$0.01-0.1$\,mas/year range. The characteristic inner jet direction
could be different within the ``core'' that remains unresolved in the VLBI images at 8~GHz. 
In the extensive 15-GHz Monitoring Of Jets in Active Galactic Nuclei (MOJAVE) survey with the VLBA, \citet{homan09} 
found that jet components often change direction after ejection. The motion of 
about half of the 203 components they studied is significantly non-radial. These misalignments are typically 
within $30\degr$ but can reach up to $80\degr$ \citep[Fig.~3 of][]{homan09}.   
Misalignments between the apparent jet position angles in compact extragalactic radio sources are well
known at larger angular scales as well. The distribution of the misalignments of the mas-scale structures
probed by VLBI, and the arcsecond-scale jets seen in the images made by the connected-element Very
Large Array (VLA) interferometer shows a bimodal form, with peaks close to $0\degr$ and $90\degr$
\citep{conway93,appl96}.


In our study, the nearly orthogonal misalignment found in 0851+202 (OJ\,287) could possibly be explained 
with a similar effect, on a smaller angular scale. 
OJ\,287 is a source known to have a
precessing jet 
\citep[][and references therein]{tate04}. Interestingly, the highest-resolution VLBI image available for
OJ\,287 \citep[86~GHz,][]{lee08} shows a hint on a $\sim$$100$~$\mu$as-scale weak extension to the compact unresolved core,
close to the direction of the apparent proper motion measured for this source ($-15.6\degr$; Table~\ref{table1}).

Proper motion directions could be more favorably compared with VLBI images made at an order of magnitude higher resolution. This can be
achieved by decreasing the observing wavelength ($\lambda$) and/or by using Space VLBI (SVLBI) to increase the length of the
interferometer baselines ($B$). (The angular resolution of an interferometer array is proportional to $\lambda/B$.) For example, a recent,
and currently the most extended, global 86-GHz VLBI survey \citep{lee08} contains single-epoch information on structural
properties of more than 100 extragalactic radio sources at sub-mas scale. A practical complication is that the extended optically thin
steep-spectrum features fade away at high frequencies. Therefore it may prove difficult to use the mm-VLBI images which are more
dominated by a single ``core'' component to define a characteristic jet direction for a large sample of objects. The technique of SVLBI
offers an increase in resolution in another way. It involves an antenna orbiting around the Earth which is co-observing with a
ground-based radio telescope network. 
The recently canceled second-generation Japanese SVLBI mission ASTRO-G \citep{tsuboi08} would have provided
ground--space baselines that exceed the Earth diameter by a factor of $\sim$$3$.

It is possible that systematic effects other than changes in the
source brightness distribution structure also affect the measured
apparent proper motions. The estimated dipole and quadrupole
harmonics in the pattern of the vector field of the proper motions
in the sky \citep[][and references therein ]{macmillan05,titov2009} could in principle be interpreted in several
ways: as a consequence of the galactocentric rotation of the Solar
system (note that our proper motion data have already been corrected for the estimated amount of 
this systematic effect); as a result of an anisotropic Hubble expansion; as an indication of the
primordial gravitational waves in the early Universe 
\citep{kristian66,pyne96,gwinn97,sovers98,macmillan05,titov2009}.  
Observationally establishing the magnitude of the 
all-sky systematic effects is complicated by the fact that the 
reference source distribution is uneven, especially in the southern 
hemisphere.  For instance, correlation coefficients between the first 
and second degree spherical harmonics could reach 0.8--0.9 due to the 
paucity of the radio sources below $-40\degr$ declination \citep{titov09}.

\citet{gontier01} suggest that at low declinations ($|\delta|
\leq 20\degr$), tropospheric mismodelling could lead to more
uncertain position estimates, preferentially in the north--south
direction. 
We could not detect any declination-dependent effect in the values of $\Delta\Phi$, 
maybe due to the small size of our sample.


In our analysis presented above, we assumed that the apparent proper motion of the sources can be represented by a linear
change
as a function of time. The VLBI position time series we used span nearly a quarter of a century for many of the objects. While
the linear approximation is valid for most of the sources, the problem in general is clearly more complex. Physically, many of
the sources 
could have undergone multiple jet component ejections over the time range studied. Although the position angle of the jet
trajectories 
(the path of the subsequent components) could remain similar for a given source, these patterns in the jet flow may change
their apparent speed.
Indeed, there are a few AGNs known to show non-linear proper motions  \citep[e.g.][]{titov07,zharov09}. 
Coordinate time series for the most extreme cases are available in the ICRF2 document \citep[Chapter 4,][]{icrf2}. 
Therefore the linear fit applied here is not necessarily the best representation of their coordinate
changes. This fact is reflected in the higher formal errors obtained for the fitted linear proper motion values for these
sources. However, here we analysed the directions, and used the fitted absolute values of the apparent proper motion only 
to select the objects that show significant motion.
Future extensions of our work could look for directional changes as well, since the apparent proper motions could follow curved
paths. It would also be interesting to find a relation between the magnitude of the detected proper motions and the structural
properties of AGNs.  

\section{Conclusions}
We studied a sample of 62 compact extragalactic radio sources for which reliable long-term VLBI time series of astrometric positions as
well as detailed 8-GHz VLBI brightness distribution models \citep{piner} were available (Table~\ref{table1}). We compared
the direction of their apparent proper motion and the characteristic direction of their $\sim$$1-10$\,mas-scale radio jet structure. 
Earlier, for a number of
individual sources, the observed astrometric proper motions have been found to be in a good agreement with the apparent motion of bright
jet components in the mas-scale radio structure \citep[e.g. 4C\,+39.25,][]{fey97,titov07}. However, the
assumption that the apparent proper motions are closely related to changes in the radio brightness distribution has not yet been tested on
a large sample of sources. With a statistical analysis, we found that there is a general correlation between the direction of
proper motion and the characteristic direction of core--jet sources. However, there are some individual objects where this relationship does not hold.
In particular, a nearly orthogonal misalignment was found in the case of OJ\,287, a blazar with a known precessing jet.  
Our results imply that the apparent proper motion is in general influenced by the changes in the source brightness
structure. 
This could be better verified in the future with higher-resolution mm-VLBI and/or Space VLBI imaging data.

\acknowledgments
We are grateful for the referee's insightful and constructive comments which certainly improved the
presentation and clarity of our work.
This work was partly supported by the Hungarian Scientific Research Fund (OTKA K72515). 
The authors thank Lajos G. Bal\'azs and Csaba Kiss for useful discussions. 
This research made use of the \anchor{http://astrogeo.org}{astrogeo.org} database of brightness distributions,
 correlated flux densities and images of compact radio sources produced with VLBI, maintained by Leonid Petrov. 
This study also made
use of the International VLBI Service for Geodesy and Astrometry (IVS)
observational data base.

\clearpage

\begin{deluxetable}{lclcrcrrcrr}                                                                                                                   
\tabletypesize{\scriptsize}                                                                                                                                                                                                                                                                          
\tablecaption{The properties of the 62 selected radio sources. \label{table1}}                                                                                  
\tablewidth{0pt}                                                                                                                                      
\tablehead{
               & \multicolumn{3}{c}{Basic properties}  &  \multicolumn{2}{c}{Structure maps} & \multicolumn{3}{c}{Geodetic VLBI observations} &  \\                                                                                                                                           
\colhead{Name} & \colhead{Class} & \colhead{$z$} &                                                                                                          
 \colhead{$D$} & \colhead{$\phi\pm\sigma_{\phi}$} & \colhead{$N^\phi_{epochs}$}                                                                              
& \colhead{$r\pm\sigma_{r}$} & \colhead{$\psi\pm\sigma_{\psi}$} & \colhead{$N^\psi_{epochs}$} & \colhead{$\Delta \Phi\pm\sigma_{\Delta \Phi}$} \\
\colhead{} & \colhead{} & \colhead{} & \colhead{} & \colhead{($\degr$)} & \colhead{} & \colhead{($\mu$as/year)} & \colhead{($\degr$)} & \colhead{} & \colhead{($\degr$)}
}                                                                                                                                                      
\startdata              
           0003$-$066 & B & 0.35 & 0 & \phantom{0}$-$68.0$\pm$\phantom{0}4.6 & 11 & \phantom{0}21.3$\pm$10.9 & \phantom{0}$-$24.3$\pm$20.3 & 1226 & \phantom{0}43.6$\pm$20.8 \\
            0104$-$408 & Q & 0.58 & 2 & \phantom{0}$+$25.8$\pm$11.4 & 7 & \phantom{00}8.7$\pm$\phantom{0}8.0 & \phantom{0}$+$51.2$\pm$53.8 & 940 & \phantom{0}25.4$\pm$55.0 \\
              0111$+$021 & B & 0.05 & 0 & $+$131.6$\pm$\phantom{0}4.3 & 8 & \phantom{0}41.0$\pm$22.5 & $+$123.3$\pm$32.8 & 178 & \phantom{00}8.3$\pm$33.1 \\
     0119$+$041 & Q(HP) & 0.64 & 0 & \phantom{0}$+$96.1$\pm$\phantom{0}2.3 & 14 & \phantom{0}14.6$\pm$\phantom{0}6.0 & \phantom{0}$-$90.2$\pm$40.6 & 1554 & \phantom{00}6.3$\pm$40.6 \\
      0119$+$115 & Q(HP) & 0.57 & 2 & \phantom{00}$+$5.9$\pm$\phantom{0}1.4 & 12 & \phantom{0}15.2$\pm$10.0 & \phantom{0}$+$18.6$\pm$24.8 & 1112 & \phantom{0}12.7$\pm$24.8 \\
     0133$+$476 & Q(HP) & 0.86 & 3 & \phantom{0}$-$28.2$\pm$\phantom{0}6.2 & 13 & \phantom{00}4.9$\pm$\phantom{0}4.6 & \phantom{0}$-$42.4$\pm$53.2 & 1466 & \phantom{0}14.2$\pm$53.6 \\
         0202$+$149 & G & 0.41 & 0 & \phantom{0}$-$51.9$\pm$\phantom{0}1.5 & 13 & \phantom{0}48.1$\pm$\phantom{0}7.4 & \phantom{0}$-$23.5$\pm$\phantom{0}6.7 & 972 & \phantom{0}28.3$\pm$\phantom{0}6.8 \\
         0229$+$131 & Q & 2.06 & 2 & \phantom{0}$+$37.5$\pm$\phantom{0}3.1 & 14 & \phantom{0}16.8$\pm$\phantom{0}5.7 & \phantom{0}$+$31.9$\pm$16.4 & 2413 & \phantom{00}5.6$\pm$16.7 \\
      0234$+$285 & Q(HP) & 1.21 & 2 & \phantom{0}$-$11.7$\pm$\phantom{0}0.5 & 13 & \phantom{0}14.6$\pm$\phantom{0}8.8 & \phantom{0}$-$55.9$\pm$45.2 & 1215 & \phantom{0}44.2$\pm$45.2 \\
      0336$-$019 & Q(HP) & 0.85 & 0 & \phantom{0}$+$69.2$\pm$\phantom{0}2.2 & 13 & \phantom{0}10.2$\pm$\phantom{0}5.3 & $-$117.1$\pm$36.0 & 1918 & \phantom{00}6.4$\pm$36.0 \\
              0402$-$362 & Q & 1.42 & 2 & \phantom{0}$+$22.6$\pm$\phantom{0}2.8 & 6 & \phantom{0}33.7$\pm$14.4 & $-$175.1$\pm$17.3 & 553 & \phantom{0}17.6$\pm$17.5 \\
           0528$+$134 & Q & 2.06 & 0 & \phantom{0}$+$45.6$\pm$12.4 & 14 & \phantom{0}14.6$\pm$\phantom{0}3.1 & \phantom{0}$+$77.9$\pm$18.4 & 3302 & \phantom{0}32.3$\pm$22.2 \\
        0537$-$441 & Q(HP) & 0.89 & 2 & \phantom{0}$+$58.1$\pm$34.8 & 8 & \phantom{0}11.3$\pm$\phantom{0}7.4 & \phantom{0}$+$11.3$\pm$30.5 & 1148 & \phantom{0}46.9$\pm$46.3 \\
           0552$+$398 & Q & 2.37 & 2 & \phantom{0}$-$71.1$\pm$\phantom{0}1.3 & 18 & \phantom{0}10.9$\pm$\phantom{0}2.8 & $+$124.0$\pm$16.5 & 4304 & \phantom{0}15.1$\pm$16.5 \\
          0642$+$449 & Q & 3.40 & 3 & \phantom{0}$+$95.0$\pm$\phantom{0}1.4 & 11 & \phantom{0}23.3$\pm$\phantom{0}5.5 & \phantom{0}$+$82.7$\pm$16.0 & 1305 & \phantom{0}12.3$\pm$16.0 \\
           0727$-$115 & Q & 1.59 & 2 & \phantom{0}$-$64.8$\pm$31.2 & 17 & \phantom{0}11.4$\pm$\phantom{0}3.1 & \phantom{0}$+$93.8$\pm$21.9 & 3457 & \phantom{0}21.4$\pm$38.1 \\
            0742$+$103 & G & 2.62 & 0 & \phantom{00}$-$7.1$\pm$\phantom{0}4.0 & 9 & \phantom{0}35.2$\pm$21.5 & \phantom{0}$-$76.4$\pm$53.8 & 325 & \phantom{0}69.3$\pm$54.0 \\
      0804$+$499 & Q(HP) & 1.43 & 3 & $+$133.8$\pm$\phantom{0}8.9 & 13 & \phantom{00}8.8$\pm$\phantom{0}6.0 & \phantom{0}$+$87.9$\pm$44.5 & 1373 & \phantom{0}45.9$\pm$45.4 \\
             0805$+$410 & Q & 1.42 & 3 & \phantom{0}$+$33.8$\pm$\phantom{0}9.1 & 3 & \phantom{0}24.1$\pm$10.8 & \phantom{0}$+$17.4$\pm$22.5 & 573 & \phantom{0}16.4$\pm$24.3 \\
          0851$+$202 & B & 0.31 & 2 & \phantom{0}$-$99.5$\pm$\phantom{0}8.3 & 15 & \phantom{0}14.5$\pm$\phantom{0}3.5 & \phantom{0}$-$15.6$\pm$10.1 & 3599 & \phantom{0}84.0$\pm$13.1 \\
           0919$-$260 & Q & 2.30 & 0 & \phantom{0}$-$86.4$\pm$\phantom{0}6.4 & 12 & \phantom{0}93.5$\pm$27.6 & \phantom{0}$-$84.5$\pm$20.4 & 409 & \phantom{00}1.9$\pm$21.4 \\
              0920$-$397 & Q & 0.59 & 2 & $-$179.8$\pm$\phantom{0}3.0 & 7 & \phantom{0}66.5$\pm$30.8 & $-$171.3$\pm$21.7 & 150 & \phantom{00}8.5$\pm$22.0 \\
         0923$+$392 & Q & 0.70 & 0 & \phantom{0}$-$80.0$\pm$\phantom{0}3.3 & 14 & \phantom{0}30.7$\pm$\phantom{0}3.1 & $+$124.7$\pm$\phantom{0}6.2 & 3898 & \phantom{0}24.7$\pm$\phantom{0}7.0 \\
             0953$+$254 & Q & 0.71 & 0 & $-$130.5$\pm$\phantom{0}0.6 & 5 & \phantom{0}30.0$\pm$\phantom{0}9.4 & \phantom{0}$+$20.1$\pm$14.4 & 974 & \phantom{0}29.3$\pm$14.4 \\
            0955$+$476 & Q & 1.87 & 3 & $+$137.7$\pm$11.3 & 14 & \phantom{00}5.3$\pm$\phantom{0}5.0 & $-$128.0$\pm$55.7 & 2033 & \phantom{0}85.7$\pm$56.8 \\
               1004$+$141 & Q & 2.71 & 0 & $+$129.9$\pm$\phantom{0}0.1 & 9 & \phantom{0}29.6$\pm$17.8 & $+$117.4$\pm$38.7 & 265 & \phantom{0}12.5$\pm$38.7 \\
         1034$-$293 & Q(HP) & 0.31 & 2 & $+$126.5$\pm$10.6 & 11 & \phantom{0}28.6$\pm$\phantom{0}6.8 & $+$137.2$\pm$13.5 & 1654 & \phantom{0}10.6$\pm$17.1 \\
           1044$+$719 & Q & 1.15 & 0 & \phantom{0}$+$66.4$\pm$20.4 & 8 & \phantom{0}27.0$\pm$\phantom{0}6.1 & \phantom{00}$-$6.7$\pm$12.4 & 1220 & \phantom{0}73.1$\pm$23.9 \\
             1101$+$384 & B & 0.03 & 2 & \phantom{0}$-$37.7$\pm$\phantom{0}3.5 & 11 & \phantom{0}36.0$\pm$15.6 & $+$154.5$\pm$24.8 & 420 & \phantom{0}12.2$\pm$25.0 \\
             1124$-$186 & Q & 1.05 & 2 & $-$148.3$\pm$24.8 & 11 & \phantom{0}10.3$\pm$\phantom{0}9.0 & $-$166.6$\pm$38.7 & 1142 & \phantom{0}18.4$\pm$46.0 \\
           1128$+$385 & Q & 1.73 & 3 & $-$163.1$\pm$\phantom{0}5.4 & 15 & \phantom{0}10.0$\pm$\phantom{0}6.1 & $+$110.7$\pm$40.1 & 1290 & \phantom{0}86.2$\pm$40.5 \\
            1145$-$071 & Q & 1.34 & 2 & \phantom{0}$-$71.6$\pm$\phantom{0}2.4 & 12 & \phantom{0}37.8$\pm$17.9 & \phantom{0}$-$25.3$\pm$24.1 & 148 & \phantom{0}46.3$\pm$24.2 \\
     1156$+$295 & Q(HP) & 0.73 & 2 & \phantom{00}$+$7.6$\pm$\phantom{0}6.9 & 12 & \phantom{0}12.7$\pm$\phantom{0}8.3 & $-$174.5$\pm$25.7 & 1289 & \phantom{00}2.2$\pm$26.6 \\
           1219$+$044 & Q & 0.97 & 3 & $+$174.5$\pm$\phantom{0}3.6 & 11 & \phantom{0}19.0$\pm$\phantom{0}6.4 & \phantom{0}$+$71.9$\pm$28.0 & 1204 & \phantom{0}77.4$\pm$28.2 \\
         1228$+$126 & G & 0.004 & 0 & \phantom{0}$-$77.6$\pm$\phantom{0}2.9 & 12 & \phantom{00}8.7$\pm$\phantom{0}8.5 & $+$174.4$\pm$44.3 & 1235 & \phantom{0}72.0$\pm$44.4 \\
         1253$-$055 & Q(HP) & 0.54 & 0 & $-$113.9$\pm$\phantom{0}1.3 & 3 & 273.4$\pm$27.8 & \phantom{0}$+$88.2$\pm$\phantom{0}9.6 & 258 & \phantom{0}22.1$\pm$\phantom{0}9.7 \\
              1255$-$316 & Q & 1.92 & 0 & \phantom{0}$+$21.9$\pm$\phantom{0}0.1 & 5 & \phantom{0}51.9$\pm$23.8 & $+$110.7$\pm$29.0 & 418 & \phantom{0}88.8$\pm$29.0 \\
              1313$-$333 & Q & 1.21 & 2 & \phantom{0}$-$85.3$\pm$18.3 & 11 & \phantom{0}29.8$\pm$22.9 & $-$141.6$\pm$44.2 & 193 & \phantom{0}56.3$\pm$47.8 \\
        1334$-$127 & Q(HP) & 0.54 & 2 & $+$153.7$\pm$\phantom{0}0.9 & 11 & \phantom{0}10.9$\pm$\phantom{0}4.8 & $-$135.8$\pm$24.8 & 2767 & \phantom{0}70.5$\pm$24.8 \\
              1351$-$018 & Q & 3.71 & 2 & $+$138.8$\pm$\phantom{0}7.9 & 6 & \phantom{0}35.8$\pm$13.4 & \phantom{0}$+$11.9$\pm$13.0 & 832 & \phantom{0}53.1$\pm$15.2 \\
             1418$+$546 & B & 0.15 & 3 & $+$130.2$\pm$\phantom{0}2.8 & 8 & \phantom{00}6.2$\pm$\phantom{0}5.9 & $-$117.2$\pm$57.4 & 817 & \phantom{0}67.5$\pm$57.5 \\
            1424$-$418 & Q(HP) & 1.52 & 2 & $+$104.5$\pm$51.3 & 7 & \phantom{0}13.8$\pm$11.2 & $+$146.3$\pm$42.7 & 792 & \phantom{0}41.9$\pm$66.7 \\
             1514$-$241 & B & 0.05 & 0 & $+$158.5$\pm$\phantom{0}1.8 & 11 & \phantom{0}40.4$\pm$36.0 & \phantom{0}$+$25.4$\pm$41.7 & 248 & \phantom{0}46.9$\pm$41.8 \\
           1606$+$106 & Q & 1.23 & 3 & \phantom{0}$-$58.9$\pm$\phantom{0}3.0 & 14 & \phantom{0}12.4$\pm$\phantom{0}5.6 & $+$175.2$\pm$18.0 & 2390 & \phantom{0}54.1$\pm$18.3 \\
          1611$+$343 & Q & 1.40 & 0 & $+$169.4$\pm$\phantom{0}2.1 & 12 & \phantom{0}25.5$\pm$\phantom{0}4.9 & $-$159.2$\pm$\phantom{0}8.4 & 1934 & \phantom{0}31.4$\pm$\phantom{0}8.7 \\
            1622$-$253 & Q & 0.79 & 2 & \phantom{0}$-$13.3$\pm$16.0 & 11 & \phantom{0}20.3$\pm$\phantom{0}8.3 & $-$153.7$\pm$18.5 & 1970 & \phantom{0}39.6$\pm$24.5 \\
       1638$+$398 & Q(HP) & 1.66 & 2 & $-$175.4$\pm$13.2 & 14 & \phantom{00}9.6$\pm$\phantom{0}9.2 & \phantom{0}$-$21.8$\pm$39.8 & 1198 & \phantom{0}26.4$\pm$41.9 \\
           1726$+$455 & Q & 0.72 & 3 & \phantom{0}$-$91.8$\pm$13.4 & 10 & \phantom{0}12.3$\pm$\phantom{0}8.6 & \phantom{0}$-$59.8$\pm$43.5 & 1302 & \phantom{0}32.0$\pm$45.5 \\
        1741$-$038 & Q(HP) & 1.05 & 2 & $-$171.6$\pm$10.3 & 13 & \phantom{00}4.4$\pm$\phantom{0}4.4 & $+$178.3$\pm$38.9 & 3551 & \phantom{0}10.1$\pm$40.2 \\
            1745$+$624 & Q & 3.89 & 3 & $-$147.2$\pm$\phantom{0}1.7 & 12 & \phantom{0}19.3$\pm$\phantom{0}8.2 & \phantom{0}$+$51.6$\pm$25.0 & 820 & \phantom{0}18.8$\pm$25.1 \\
      1749$+$096 & Q(HP) & 0.32 & 2 & \phantom{0}$+$32.3$\pm$\phantom{0}8.8 & 17 & \phantom{00}5.1$\pm$\phantom{0}4.7 & $-$166.5$\pm$37.1 & 2792 & \phantom{0}18.7$\pm$38.2 \\
     1803$+$784 & Q(HP) & 0.68 & 2 & \phantom{0}$-$99.1$\pm$\phantom{0}2.8 & 12 & \phantom{00}9.9$\pm$\phantom{0}3.4 & \phantom{0}$-$79.3$\pm$20.5 & 2449 & \phantom{0}19.8$\pm$20.7 \\
             1908$-$201 & Q & 1.12 & 2 & \phantom{0}$+$33.1$\pm$\phantom{0}1.4 & 11 & \phantom{0}19.8$\pm$18.7 & $+$164.8$\pm$34.9 & 752 & \phantom{0}48.3$\pm$34.9 \\
      1921$-$293 & Q(HP) & 0.35 & 2 & \phantom{0}$+$27.4$\pm$\phantom{0}5.1 & 14 & \phantom{0}22.6$\pm$\phantom{0}7.6 & \phantom{0}$-$27.5$\pm$16.1 & 1743 & \phantom{0}54.9$\pm$16.9 \\
         1958$-$179 & Q(HP) & 0.65 & 2 & $-$133.6$\pm$36.9 & 3 & \phantom{0}18.3$\pm$\phantom{0}6.5 & \phantom{0}$+$93.3$\pm$20.2 & 1341 & \phantom{0}46.9$\pm$42.1 \\
              2136$+$141 & Q & 2.43 & 3 & $+$119.2$\pm$13.5 & 8 & \phantom{0}16.2$\pm$\phantom{0}8.1 & $-$122.2$\pm$32.1 & 1031 & \phantom{0}61.4$\pm$34.8 \\
           2145$+$067 & Q & 0.99 & 1 & $+$130.0$\pm$\phantom{0}3.1 & 19 & \phantom{0}26.4$\pm$\phantom{0}5.7 & \phantom{0}$-$29.1$\pm$10.6 & 2217 & \phantom{0}20.9$\pm$11.1 \\
           2200$+$420 & B & 0.07 & 0 & $-$169.1$\pm$\phantom{0}5.2 & 12 & \phantom{0}20.4$\pm$\phantom{0}6.6 & \phantom{0}$-$16.1$\pm$15.8 & 1026 & \phantom{0}27.0$\pm$16.6 \\
           2230$+$114 & Q(HP) & 1.04 & 0 & $+$157.8$\pm$\phantom{0}1.4 & 6 & \phantom{0}29.4$\pm$13.1 & $+$129.7$\pm$26.8 & 175 & \phantom{0}28.1$\pm$26.8 \\
        2234$+$282 & Q(HP) & 0.80 & 0 & $-$132.2$\pm$11.2 & 12 & \phantom{0}31.0$\pm$\phantom{0}4.2 & $-$112.0$\pm$\phantom{0}9.0 & 2271 & \phantom{0}20.1$\pm$14.4 \\
       2243$-$123 & Q(HP) & 0.63 & 0 & \phantom{00}$+$1.7$\pm$\phantom{0}6.4 & 12 & \phantom{0}17.5$\pm$\phantom{0}7.5 & $+$119.3$\pm$32.2 & 971 & \phantom{0}62.4$\pm$32.9 \\
             2255$-$282 & Q & 0.93 & 2 & $-$136.4$\pm$\phantom{0}2.1 & 9 & \phantom{0}17.5$\pm$10.0 & $-$177.8$\pm$23.8 & 1398 & \phantom{0}41.4$\pm$23.9 \\
\enddata                                                                                                                                         
 \tablecomments{Col.(1): Name of the radio source derived from B1950 equatorial coordinates.                                                                                                              
 Col.(2): Optical class of the source. B = BL Lac object, Q = quasar, G = galaxy, HP refers
to high polarization. The class information was taken from
\citet{veron03}.                                                                                                                              
 Col.(3): Redshift values \citep[adopted from][]{piner}.                                                                                                                                
 Col.(4): Shows whether the specific object is a defining source of the ICRF1/ICRF2. $D$=1 and 2
mark ICRF1 and ICRF2 defining sources, respectively. $D$=3 indicates sources that are defining objects 
in both ICRF1 and ICRF2.                             
 Col.(5): Position angle of the final characteristic core--jet direction
($\phi$) and its uncertainty ($\sigma_{\phi}$). Position angle is measured in degrees from north through east.                                                                            
 Col.(6): Number of observations used to compute $\phi$.                                                                               
 Col.(7): Radial component of the apparent proper motion ($r$) and its uncertainty ($\sigma_{r}$) in $\mu$as/year.                                                                                                                           
 Col.(8): Position angle of the apparent proper motion ($\psi$) and its uncertainty ($\sigma_{\psi}$) in degrees.                                                                                                              
 Col.(9): Number of observations used to compute $\psi$.
 Col.(10): The smallest angle between the direction of the apparent proper motion and the characteristic core--jet direction in degrees.                                                                                                                                                                           
}                                                                                                                                                    
\end{deluxetable} 

\clearpage

\clearpage

\clearpage

\end{document}